\RequirePackage{fix-cm}
\documentclass{svjour3}                     % onecolumn (standard format)
\smartqed  % flush right qed marks, e.g. at end of proof
\usepackage{graphicx}
\usepackage{bm}
\usepackage{mathptmx}      % use Times fonts if available on your TeX system
\usepackage{latexsym}
\begin{document}

\title{Superconductivity by Berry connection from many-body wave functions: revisit to Andreev$-$Saint-James reflection and
Josephson effect%\thanks{Grants or other notes
%about the article that should go on the front page should be
%placed here. General acknowledgments should be placed at the end of the article.}
}
%\subtitle{Do you have a subtitle?\\ If so, write it here}

%\titlerunning{Short form of title}        % if too long for running head

\author{Hiroyasu Koizumi  %etc.
}

%\authorrunning{Short form of author list} % if too long for running head

\institute{Hiroyasu Koizumi  \at
              Division of Quantum Condensed Matter Physics, Center for Computational Sciences, University of Tsukuba,Tsukuba, Ibaraki 305-8577, Japan \\
              Tel.: +81-29-8536403\\
              Fax: +81-29-8536403\\
              \email{koizumi.hiroyasu.fn@u.tsukuba.ac.jp}           %  \\
%             \emph{Present address:} of F. Author  %  if needed
}

\date{Received: date / Accepted: date}
% The correct dates will be entered by the editor

\maketitle

\begin{abstract}
Although the standard theory of superconductivity based on the BCS theory is a successful one, several experimental results indicate the necessity for a fundamental revision. 

We argue that the revision is on the origin of the phase variable for superconductivity; this phase appears as a consequence of the electron-pairing in the standard theory, but its origin is a Berry connection arising from many-body wave functions. 
 When this Berry connection is non-trivial, it gives rise to a collective mode that generates supercurrent; this collective mode creates number-changing operators for particles participating in this mode, and these number-changing operators stabilize the superconducting state by exploiting the Cooper instability.

In the new theory, the role of the electron-pairing is to stabilize the nontrivial Berry connection; it is not the cause of superconductivity.
 In BCS superconductors, however, the simultaneous appearance of the nontrivial Berry connection
and the electron-pairing occurs. Therefore, the electron-pairing amplitude can be used as an order parameter for the superconducting state.

We revisit the Andreev$-$Saint-James reflection and the Josephson effect. They are explained as consequences of the presence of the
Berry connection. Bogoliubov quasiparticles are replaced by the particle-number conserving 
Bogoliubov excitations that describe the transfer of electrons between the collective and single 
particle modes. There are two distinct cases for the Josephson effect; one of them contains the common Bogoliubov excitations for the two superconductors in the junction, and the other does different Bogoliubov excitations for different superconductors. The latter case is the one considered in the standard theory; in this case, the Cooper pairs tunnel through without Bogoliubov excitations, creating an impression that the supercurrent is a flow of Cooper pairs; however, it does not explain the observed ac Josephson effect under the experimental boundary condition. On the other hand, 
 the former case explains the ac Josephson effect under the experimental boundary condition.
 In this case, it is clearly shown that the supercurrent is a flow of electrons brought about by the non-trivial Berry connection which provides an additional $U(1)$ gauge field besides the electromagnetic one.

% \PACS{PACS code1 \and PACS code2 \and more}
% \subclass{MSC code1 \and MSC code2 \and more}
\end{abstract}

\section{Introduction}

The current standard theory of superconductivity is the one based on the BCS theory \cite{BCS1957}.
In this theory, the order parameter of superconducting state is the electron-pairing amplitude or the pair potential. This gives rise to an energy gap for single-particle excitations and provides rigidity of the superconducting state against perturbations.
The standard theory has been successfully applied to many superconducting materials; a notable point is
that it successfully explains the superconducting transition temperatures for superconductors whose normal states are simple metals.

In 1986, high temperature superconductivity was found in ceramics \cite{Muller1986}. Since then, a great deal of effort has been put into the elucidation of the mechanism of it.
In spite of all the efforts, the widely-accepted theory has not been obtained, yet.
A notable point of the cuprate superconductivity is that the superconducting transition temperature for the optimally-doped sample is not the pairing energy gap formation temperature, but the stabilization temperature for loop currents of the superconducting coherence length size \cite{Kivelson95}. 

Efforts to elucidate the cuprate superconductivity have lead some people to reexamine the theory of superconductivity from fundamental levels \cite{Hirsch2009,Koizumi2011}.
Then, it is noticed that there are two solid experimental facts that point to the need for fundamental revisions in the standard theory \cite{Hirsch2009,Koizumi2011}.

One of them is the reversible phase transitions between normal and superconducting phases in the $H$-$T$ plane (for Type I superconductors)\cite{Hirsch2017,Hirsch2018,Hirsch2020,koizumi2020b}. 
A series of work \cite{Keesom1934a,Keesom1934b,Keesom1937,Keesom} indicate that the superconducting-normal state transition in the presence of a magnetic field occurs without energy dissipation, and the state of the art calorimetry indicates that 99.99\% of the supercurrent stops without current carriers undergoing irreversible collisions (see Appendix B of Ref.~\cite{Hirsch2017}). 

 However, such a transition is impossible in the standard theory; according to the standard theory, paired electrons flow without dissipation but single electrons flow with dissipation, thus, the supercurrent generated by the flow of electron pairs in the magnetic field inevitably produces the Joule heat during the superconducting to normal phase transition due to the existence of a significant number of broken pairs that flow with dissipation. 
 
The other is the mass of the electron in the London moment \cite{Hirsch2013b,koizumi2021}. 
Inside a rotating superconductor, a magnetic field is created by the supercurrent produced in the surface region.
The London moment is the magnetic moment produced by this supercurrent \cite{London1950}. 
The London moment has been measured many times using different materials, ranging from the conventional superconductor \cite{Hildebrandt1964,Zimmerman1965,Brickman1969,Tate1989,Tate1990} to the high T$_{\rm c}$ cuprates \cite{VERHEIJEN1990a,Verheijen1990} and heavy fermion superconductors \cite{Sanzari1996}. The results always indicate that the mass $m$ is the free electron mass $m_e$ if the electron charge $q=-e$ is employed. However,
the standard theory predicts it to be an effective mass $m^{\ast}$, contradicting the experimental results.

The resolution for the above two discrepancies is provided using a new theory of superfluid that attributes the superfluidity to the appearance of  the nontrivial Berry connection from many-body wave functions \cite{koizumi2019,koizumi2021}. 
In this theory, the supercurrent is explained as a topologically protected current generated by the collective mode created by the nontrivial Berry connection.

A salient feature of the new theory is that it is formulated in a particle number conserving way. In the standard theory, however, the particle number non-conserving state vector is used, and the use of it gives rise to the phase variable that explains the Meissner effect and supercurrent generation. Bogoliubov quasiparticles appear in the standard theory are superpositions of electrons and holes, which can only be meaningful in the particle number non-conserving formalism. In the new theory, however, the Bogoliubov quasiparticles are replaced by excitations that describe the transfer of electrons between the collective and single-particle modes with keeping the particle number fixed.

The Andreev$-$Saint-James (ASJ) reflection \cite{Andreev1964,Saint-James} and the Josephson effect \cite{Josephson62} are phenomena,
where the phase variable of superconductivity plays an essential role, and explained
using the particle number non-conserving formalism in the standard theory.  
A purpose of the present work is to provide explanations for them
using  the particle number conserving formalism of the new theory, where the phase variable 
is identified as a Berry phase.
Those electrons whose coordinates are the arguments of the same wave function, actually, interact through the gauge field created by the wave function they share \cite{koizumi2019}.
This gauge field is calculated as a Berry connection.
 It is a $U(1)$ gauge field, just like the $U(1)$ gauge field of the electromagnetism. Thus, there are two  $U(1)$ gauge fields in the system.

We will also show that the presence of the Berry connection modifies Maxwell's equations, and the Lorentz interaction between charged particles and electromagnetic field. Actually, the modified Lorentz interaction gives rise to Aharonov-Bohm type effects \cite{AB1959} that cannot be described by the Lorentz force. This modification affects the magnetic energy part of the phenomenological Ginzburg-Landau theory \cite{GL} in such a way that Abrikosov's vortices \cite{Abrikosov} appear, naturally.
 Then, the superconducting coherence length is regard as the core size of the loop currents generated by the Berry connection that exist
 even without  applied magnetic field. 

 The organization of the present work is as follows;
Supercurrent flow in the new and standard theories are compared in Section~\ref{sec3.5}. 
  In Section~\ref{sec4}, the Andreev$-$Saint-James reflection is revisited. In Section~\ref{sec5}, the Josephson effect is revised. 
 The conclusion of the present work is presented in Section~\ref{sec6}.
 The modification of Maxwell's equations in the presence of the Berry connection from many-body wave functions, and its consequences in the Ginzburg-Landau theory is given in Appendices, to preserve the flow of the explanation on the Andreev$-$Saint-James reflection and the Josephson effect.
 
 \section{Supercurrent flow}
\label{sec3.5}

Let us compare the supercurrent in the new theory and that in the standard theory.

In the standard theory, the supercurrent is a flow of Cooper pairs.
 In the original BCS theory, the normal metallic state is assumed to be well-described by the free electrons with the effective mass $m^{\ast}$.
 Then, the electron field operators are given by
\begin{eqnarray}
\hat{\Psi}_{\sigma}({\bf r})={ 1 \over \sqrt{\cal V}}\sum_{\bf k} e^{i {\bf k} \cdot {\bf r}} c_{{\bf k} \sigma}
\label{f1}
\end{eqnarray}
where ${\cal V}$ is the system volume, ${\bf k}$ and ${\bf r}$ are the wave vector and coordinate of the electron, respectively, and
$c_{{\bf k} \sigma}$ is the annihilation operator for the electron with the wave vector ${\bf k}$ and spin$\sigma$.

The electron-pairing amplitude is expressed as
\begin{eqnarray}
\langle \hat{\Psi}_{\downarrow}({\bf r}_2) \hat{\Psi}_{\uparrow}({\bf r}_1) \rangle=\sum_{\bf q} e^{i {\bf q} \cdot {\bf R}}\Delta_{\bf q} ({\bf r})
\label{pair-pot}
\end{eqnarray}
where $\langle \hat{O}\rangle$ denotes the expectation value of the operator $\hat{O}$, and
 \begin{eqnarray}
\Delta_{\bf q}({\bf r})=
\sum_{{\bf p}}' e^{i {\bf p} \cdot {\bf r}}
\langle c_{-{\bf p}+{{\bf q} \over 2} \downarrow}  c_{{\bf p}+{{\bf q} \over 2}\uparrow}\rangle
\end{eqnarray}
is the electron-pairing amplitude with momentum ${\bf q}$, ${\bf r}={\bf r}_1 -{\bf r}_2$ is the relative position vector between the pairing electrons,  ${\bf R}={1 \over 2}({\bf r}_1+{\bf r}_2)$ is the center-of-mass position vector of the pairing electron, and the sum is taken over ${\bf p}$ near the Fermi level where that the attractive interaction between electrons exists.

Using $\Delta_{\bf q}({\bf r})$ the macroscopic wave function in the standard theory is given by
  \begin{eqnarray}
\Psi_{\rm GL}({\bf R})=gC\sum_{{\bf q}} e^{i {\bf q}\cdot {\bf R}}\Delta_{\bf q}(0)
\end{eqnarray}
where $g$ is the parameter for the attractive electron-electron interaction, and $C$ is a constant.

Considering the phase change of $\langle c_{-{\bf p}+{{\bf q} \over 2} \downarrow}  c_{{\bf p}+{{\bf q} \over 2}\uparrow}\rangle$ by the gauge transformation, $\nabla_{\bf R} \Psi_{\rm GL}({\bf R})$ should be modified in the presence of the magnetic field as
  \begin{eqnarray}
\nabla_{\bf R} \Psi_{\rm GL}({\bf R})\rightarrow \left[  \nabla_{\bf R} +{{2ei} \over {\hbar} } {\bf A}^{\rm em}\right] \Psi_{\rm GL}({\bf R})
\end{eqnarray}

This yields the kinetic energy term in Eq.~(\ref{mat2}) in the Ginzburg-Landau theory.
Since the kinetic mass for the electrons in the Cooper pair is the effective mass $m^{\ast}$, $m$ in it is $m=2m^{\ast}$, where
$2$ comes from the electron-pair.
  
Now we consider the same problem using the new theory. This problem is dealt with our recent publication \cite{koizumi2021}, but  we reproduce it here succinctly in the following for the later use.

 We use the coordinate dependent basis functions $u_{n}({\bf r}), v_{n}({\bf r})$ that are different from plane waves.
 Then, the field operators become
\begin{eqnarray}
\hat{\Psi}_{\uparrow}({\bf r})&=&\sum_{n} e^{-{i \over 2}\hat{\chi} ({\bf r})}\left( \gamma_{{n} \uparrow } u_{n}({\bf r})  -\gamma^{\dagger}_{{n} \downarrow } v^{\ast}_{n}({\bf r}) \right)
\nonumber
\\
\hat{\Psi}_{\downarrow}({\bf r})&=&
\sum_{n} e^{-{i \over 2}\hat{\chi} ({\bf r})} \left( \gamma_{{n} \downarrow } u_{n}({\bf r}) +\gamma^{\dagger}_{{n} \uparrow } v^{\ast}_{n}({\bf r}) \right)
\label{f2}
\end{eqnarray}

The particle number conserving Bogoliubov operators satisfy
\begin{eqnarray}
\gamma_{n \sigma}|{\rm Gnd}(N) \rangle=0
\end{eqnarray}
where $N$ is the total number of particles.

The ground state satisfies on the operation of  $e^{-{i \over 2} \hat{\chi}({\bf r})}$ as
\begin{eqnarray}
e^{-{i \over 2} \hat{\chi}({\bf r})}|{\rm Gnd}(N) \rangle= e^{-{i \over 2} {\chi}({\bf r})}|{\rm Gnd}(N-1) \rangle
\label{eqBC}
\end{eqnarray}
The operator $e^{-{i \over 2} \hat{\chi}({\bf r})}$ is the number changing operator that removes one electron from the collective modes at the position ${\bf r}$.
This also means that the Bogoliubov operators in Eq.~(\ref{f2}) conserve the particle number. We call them ``the particle-number conserving Bogoliubov operators''. They describe the transfer of electrons between the collective mode described by $\chi$ and single-particle mode
(this point may be seen in Eq.~(\ref{Bog}) in the next section).

The phase factor $e^{-{i \over 2} {\chi}({\bf r})}$ arises due to the presence of the Berry connection. The connection of geometry embedded 
in the superconducting state is manifested as the appearance of the phase factor given by
\begin{eqnarray}
\langle e^{{i \over 2} \hat{\chi}({\bf r}_2)}e^{-{i \over 2} \hat{\chi}({\bf r}_1)} \rangle= e^{{i \over 2} \int_{{\bf r}_1}^{{\bf r}_2} \nabla \chi({\bf r})\cdot d{\bf r}}
\end{eqnarray}

Now, let us consider the electronic Hamiltonian expressed as
\begin{eqnarray}
H=\sum_{\sigma} \int d^3 r \hat{\Psi}^{\dagger}_{\sigma}({\bf r}) h({\bf r}) \hat{\Psi}_{\sigma}({\bf r}) 
-{1 \over 2} \sum_{\sigma, \sigma'}\int d^3 r d^3 r' V_{\rm eff}({\bf r}, {\bf r}') \hat{\Psi}^{\dagger}_{\sigma}({\bf r}) \hat{\Psi}^{\dagger}_{\sigma'}({\bf r}') \hat{\Psi}_{\sigma'}({\bf r}') \hat{\Psi}_{\sigma}({\bf r}) 
\nonumber
\\
\end{eqnarray}
where $-V_{\rm eff}$ is the effective interaction between electrons, and $h({\bf r})$ is the single-particle Hamiltonian given by
\begin{eqnarray}
h({\bf r})={ 1 \over {2m_e}} \left( { \hbar \over i} \nabla +{e \over c} {\bf A}^{\rm em} \right)^2+U({\bf r})-\mu 
\end{eqnarray}
with $U({\bf r})$ being a potential energy and $\mu$ chemical potential.

We perform the mean field approximation 
\begin{eqnarray}
H^{\rm MF}&=&\sum_{\sigma} \int d^3 r \hat{\Psi}^{\dagger}_{\sigma}({\bf r}) h({\bf r}) \hat{\Psi}_{\sigma}({\bf r}) 
+\int d^3 r d^3 r' 
\left[ \Delta({\bf r}, {\bf r}')\hat{\Psi}^{\dagger}_{\uparrow}({\bf r}) \hat{\Psi}^{\dagger}_{\downarrow}({\bf r}') e^{-{i \over 2}(\hat{\chi}({\bf r}) +\hat{\chi}({\bf r}')) }
+{\rm H. c.} \right]
\nonumber
\\
&+&\int d^3 r d^3 r' 
{ {|\Delta({\bf r}, {\bf r}')|^2} \over {V_{\rm eff}({\bf r}, {\bf r}') }}
\nonumber
\\
\end{eqnarray}
where the gap function $\Delta({\bf r}, {\bf r}')$ is defined as 
\begin{eqnarray}
 \Delta({\bf r}, {\bf r}')= V_{\rm eff}({\bf r}, {\bf r}')\langle e^{{i \over 2}(\hat{\chi}({\bf r}) +\hat{\chi}({\bf r}')) }
\hat{\Psi}_{\uparrow}({\bf r}) \hat{\Psi}_{\downarrow} ({\bf r'}) \rangle
\label{pair-pot2}
\end{eqnarray}

Due to the factor $ e^{{i \over 2}(\hat{\chi}({\bf r}) +\hat{\chi}({\bf r}'))}$ that increase the number of electrons by two, the expectation value is calculated using the particle number fixed state in contrast to the standard theory.

Using the relation in Eq.~(\ref{eqBC}) and requiring  $H^{\rm MF}$ to become
\begin{eqnarray}
H^{\rm MF}=\sum_{n, \sigma} \epsilon_n \gamma_{n \sigma}^{\dagger}\gamma_{n \sigma}+E_{\rm const}
\end{eqnarray}
where $E_{\rm const}$ is a constant, 
 the following system of equations is obtained,
\begin{eqnarray}
\epsilon_n u_n({\bf r})&=&
\bar{h}({\bf r}) u_n({\bf r})+\int d^3 r'\Delta ({\bf r},{\bf r}')v_n({\bf r}')
\nonumber
\\
\epsilon_n v_n({\bf r})&=&-
 \bar{h}^{\ast}({\bf r}) v_n({\bf r})+\int d^3 r'\Delta^{\ast}({\bf r},{\bf r}')u_n({\bf r}')
 \label{e1}
\end{eqnarray}
where 
\begin{eqnarray}
\bar{h}({\bf r})={ 1 \over {2m_e}} \left( { \hbar \over i} \nabla +{e \over c} {\bf A}^{\rm em}-{ \hbar \over 2} \nabla \chi \right)^2+U({\bf r})-\mu 
 \label{e2}
\end{eqnarray}
and 
\begin{eqnarray}
\Delta({\bf r}, {\bf r}')=V_{\rm eff}({\bf r}, {\bf r}')\sum_n \left[ u_n({\bf r}) v^{\ast}_n({\bf r}')(1- f(\epsilon_n))-u_n({\bf r}') v^{\ast}_n({\bf r})f(\epsilon_n) \right]
 \label{e3}
\end{eqnarray}
with $f(\epsilon_n)$ being the Fermi function at temperature T ($k_B$ is the Boltzmann constant),
\begin{eqnarray}
f(\epsilon_n)=(e^{ {\epsilon_n} \over {k_{B} T}} +1)^{-1}
\end{eqnarray}

The above system of equation is the particle number conserving version of the Bogoliubov-de~Gennes equations \cite{deGennes,koizumi2019}.

Note that the gauge potential in the single particle Hamiltonian $\bar{h}({\bf r})$ is the effective one,
\begin{eqnarray}
{\bf A}^{\rm eff}={\bf A}^{\rm em}-{ {\hbar c} \over {2e}} \nabla \chi
\end{eqnarray}
 
If we solve the system of equations composed of Eqs.~(\ref{e1}), (\ref{e2}), and (\ref{e3}) with the condition ${\bf A}^{\rm em}-{ {\hbar c}\over {2e}} \nabla \chi=0$, we obtain the currentless solutions
for $u_n, v_n$, which we denote as $\acute{u}_n, \acute{v}_n$. 

We may construct ${u}_n, {v}_n$ using $\acute{u}_n, \acute{v}_n$ as follows,
\begin{eqnarray}
u_n({\bf r})\approx \acute{u}_n ({\bf r})e^{{i \over 2}{\chi}({\bf r})}, \quad v_n({\bf r})=\acute{v}_n ({\bf r})e^{-{i \over 2}{\chi}({\bf r})}
\label{equv}
\end{eqnarray}
where  $\nabla \chi$ is obtained by the requirement of the conservation of local charge, and single-valuedness of the wave function with respect to electron coordinates \cite{koizumi2019,koizumi2020c}.

\section{Andreev$-$Saint-James reflection}
\label{sec4}

In the Andreev$-$Saint-James reflection (we call it just the Andreev reflection, below), the phase factor of the pair potential plays the central role.
We examine how this phase factor arises from the Berry connection of the new theory. 

Let us consider the hopping Hamiltonian between the normal state and the superconducting state equipped with the Berry connection,
  \begin{eqnarray}
H_{\rm Super-Metal}=-t\sum_{{\bf k}, \sigma} \left(  c^{\dagger}_{i \sigma} M_{{\bf k} \sigma}+M^{\dagger}_{{\bf k} \sigma} c_{ i \sigma}  \right)
\end{eqnarray}
where $M_{{\bf k} \sigma}^{\dagger}$ and $M_{{\bf k} \sigma}$ are creation and annihilation operators for the electrons in the normal metal part, respectively.  $c_{ i \sigma}^{\dagger}$ and $c_{ i \sigma}$ are creation and annihilation operators for the electron in the superconductor part at the
interface site $i$, respectively.

Using the number-changing operators and the Bogoliubov operators defined in the previous section, annihilation and creation operators for the electrons in the superconducting state are given by
 \begin{eqnarray}
 c_{ i \sigma} &=&\sum_{n}[ u^{n}_{i}\gamma_{n \sigma}-\sigma (v^{n}_{i})^{\ast}\gamma_{n -\sigma}^{\dagger}] e^{ -{i \over 2} \hat{\chi}_i}
 \nonumber
 \\
 c^{\dagger}_{ i \sigma} &=&\sum_{n}[ (u^{n}_{i \sigma})^{\ast}\gamma^{\dagger}_{n \sigma}-\sigma v^{n}_{i}\gamma_{n -\sigma}] e^{{i \over 2} \hat{\chi}_i}
 \label{Bog}
  \end{eqnarray}
  where $\sigma=1$ for up-spin state, and $\sigma=-1$ for down-spin state.

Cooper-pair annihilation and creation operators are given by
   \begin{eqnarray}
 c_{ i \uparrow}  c_{ i \downarrow} &=&\sum_{n, n'}[ u^{n}_{i}\gamma_{n \uparrow}- (v^{n}_{i})^{\ast}\gamma_{n \downarrow}^{\dagger}] e^{ -{i \over 2} \hat{\chi}_i} [u^{n'}_{i}\gamma_{n' \downarrow}+(v^{n'}_{i})^{\ast}\gamma_{n' \uparrow}^{\dagger}] e^{ -{i \over 2} \hat{\chi}_i}
 \nonumber
 \\
  &\approx&\sum_{n} u^{n}_{i}(v^{n}_{i})^{\ast} \left[ \langle \gamma_{n \uparrow}\gamma_{n \uparrow}^{\dagger}\rangle 
  -\langle \gamma_{n \downarrow}^{\dagger}\gamma_{n \downarrow}\rangle\right]e^{- {i} \hat{\chi}_i}
  \nonumber
  \\
  &=&e^{- {i} \hat{\chi}_i}\sum_{n}u^{n}_{i}(v^{n}_{i})^{\ast}[1 -2f(\epsilon_n)]
   \nonumber
 \\
 c^{\dagger}_{ i \downarrow}  c^{\dagger}_{ i \uparrow}  &\approx&
 e^{ {i} \hat{\chi}_i}\sum_{n}(u^{n}_{i})^{\ast}v^{n}_{i}[1 -2f(\epsilon_n)]
\end{eqnarray}
where the products of Bogoliubov operators are replaced by their expectation values.
The above shows that Cooper-pairs can move without Bogoliubov excitations. This feature makes an impression that supercurrent is a flow of 
Cooper-pairs in the standard theory.

Now, let us derive the term for the Andreev reflection by applying the second order perturbation theory using $H_{\rm Super-Metal}$,
\begin{eqnarray}
&&\langle H_{\rm Super-Metal}{1 \over {E_0 - H_0}} H_{\rm Super-Metal} \rangle_{\rm Bog}
\nonumber
\\
& \approx & t^2 
\left\langle \sum_{{\bf k}, {\bf k}', \sigma, \sigma'}
\left[ c^{\dagger}_{i \sigma} M_{{\bf k} \sigma}{1 \over {E_0 - H_0}} c^{\dagger}_{i \sigma'} M_{{\bf k}' \sigma'} 
+
M^{\dagger}_{{\bf k} \sigma} c_{ i \sigma}  {1 \over {E_0 - H_0}} M^{\dagger}_{{\bf k}', \sigma'} c_{ i \sigma'}  
\right]
\right\rangle_{\rm Bog}
\nonumber
\\
&\approx & -
 \sum_{{\bf k}, {\bf k}', n}{{2t^2} \over {\epsilon_n}}
\left[ (u^{n}_{i})^{\ast}
v^{n}_{i} M_{{\bf k} \uparrow} M_{{\bf k}' \downarrow} e^{ {i} \hat{\chi}_i}
+
(v^{n}_{i})^{\ast}
u^{n}_{i} M^{\dagger}_{{\bf k}' \downarrow} M^{\dagger}_{{\bf k} \uparrow}  e^{ -{i} \hat{\chi}_i}
\right]
\label{Andreev1}
\end{eqnarray}
where 
$ \langle  \cdots \rangle_{\rm Bog}$ denotes that the expectation value is calculated for the products of Bogoliubov operators.

The above effective Hamiltonian indicates that when an electron in the normal state is reflected back as a hole, the phase factor $e^{ {i}{\chi}_i}$ is attached,
and a hole in the normal state is reflected back as an electron, the phase factor $e^{ -{i}{\chi}_i}$ is attached; they are salient features of the Andreev reflection.
In the following, we outline how the Andreev equations are derived from the particle-number conserving Bogoliubov-de Gennes method by following Ref. \cite{Zhu2016}. 

From Eq.~(\ref{e1}) using (\ref{equv}), the particle-number conserving Bogoliubov-de Gennes equations are expressed as
\begin{eqnarray}
\epsilon_n \acute{u}_n({\bf r})&=&
{h}({\bf r}) \acute{u}_n({\bf r})+\int d^3 r' e^{-{i \over 2}(\chi({\bf r})+\chi({\bf r}'))}\Delta ({\bf r},{\bf r}')\acute{v}_n({\bf r}')
\nonumber
\\
\epsilon_n \acute{v}_n({\bf r})&=&-{h}^{\ast}({\bf r}) \acute{v}_n({\bf r})+\int d^3 r' e^{{i \over 2}(\chi({\bf r})+\chi({\bf r}'))}\Delta^{\ast}({\bf r},{\bf r}')\acute{u}_n({\bf r}')
 \label{e11}
\end{eqnarray}
with $h({\bf r})$ as the single-particle Hamiltonian instead of $\bar{h}({\bf r})$. In order to obtain the Andreev equations, we need to have the same single-particle Hamiltonian in both the normal metallic and superconducting parts, thus, we use  $h({\bf r})$.

Then, we may regard 
\begin{eqnarray}
\bar{\Delta}\left({{{\bf r}+{\bf r}'} \over 2}, {\bf r}-{\bf r}' \right)=e^{-{i \over 2}(\chi({\bf r})+\chi({\bf r}'))}\Delta ({\bf r},{\bf r}')
\end{eqnarray}
 as the pair potential for the Andreev reflection. 
 Due to the phase factor $e^{-{i \over 2}(\chi({\bf r})+\chi({\bf r}'))}$ in $\bar{\Delta}$, it describes the phase in 
the Andreev reflection.
 
 By taking the Fourier transformation of $\bar{\Delta}\left({{{\bf r}+{\bf r}'} \over 2}, {\bf r}-{\bf r}' \right)$ with respect to the relative coordinate ${\bf s}={\bf r}-{\bf r}'$, and denoting the center-of-mass coordinate as ${\bf r}$, the following pair potential is obtained,
 \begin{eqnarray}
\bar{\Delta}({\bf r}, {\bf k})=\int d^3 s \bar{\Delta}({\bf r}, {\bf s})e^{-i{\bf k}\cdot {\bf s}}
\end{eqnarray}

In the weak coupling case, only the wave vectors very close to the Fermi surface are important. Then, ${\bf k}$ dependence of $\bar{\Delta}({\bf r}, {\bf k})$ comes only from the direction of the Fermi wave vector ${\bf k}_F$, or the unit vector on the Fermi surface ${\bf e}_{{\bf k}_F}$. Thus, $\bar{\Delta}({\bf r}, {\bf k})$
can be approximated as 
 \begin{eqnarray}
\bar{\Delta}_{{\bf e}_{{\bf k}_F}}({\bf r}) \approx \bar{\Delta}({\bf r}, {\bf k}_F) 
\end{eqnarray}
 
We also separate the fast Fermi wave vector oscillation from $\acute{u}_n({\bf r}), \acute{v}_n({\bf r})$  using $\bar{u}_n({\bf r}), \bar{v}_n({\bf r})$
defined as 
 \begin{eqnarray}
\bar{u}_n({\bf r})=e^{- i{\bf k}_F \cdot {\bf r}} \acute{u}_n({\bf r}), \quad  \bar{v}_n({\bf r})=e^{- i{\bf k}_F \cdot {\bf r}} \acute{v}_n({\bf r})
\end{eqnarray}

By considering the case with $\xi_{\rm BCS} \gg k_F^{-1}$,
Eq.~(\ref{e11}) becomes the following Andreev equations
\begin{eqnarray}
\epsilon_n \bar{u}_n({\bf r})&=& -i {\hbar^2 \over m_e} {\bf k}_F \cdot \nabla \bar{u}_n({\bf r})
+\bar{\Delta}_{ {\bf e}_{{\bf k}_F}} ({\bf r}) \bar{v}_n({\bf r})
\nonumber
\\
\epsilon_n \bar{v}_n({\bf r})&=& i {\hbar^2 \over m_e} {\bf k}_F \cdot \nabla \bar{v}_n({\bf r})
+\bar{\Delta}^{\ast}_{ {\bf e}_{{\bf k}_F}} ({\bf r}) \bar{u}_n({\bf r})
 \label{e12}
\end{eqnarray}
The effect of the phase factor $e^{{i}{\chi}_i}$ is included in the phase of $\bar{\Delta}_{ {\bf e}_{{\bf k}_F}} ({\bf r})$ at the reflection point ${\bf r}={\bf r}_i$. 

%In the standard theory, the above corresponds to
%\begin{eqnarray}
% \Delta({\bf r})\approx g e^{-i{\chi}({\bf r})} \langle 
%\hat{\Psi}_{\uparrow}({\bf r}) \hat{\Psi}_{\downarrow} ({\bf r'}) \rangle_{\rm BCS}
%\end{eqnarray}
%where $ \langle  \cdots \rangle_{\rm BCS}$ denotes that the expectation value is calculated using the particle number non-conserving state vector of the BCS theory.
%The supercurrent is due to this $e^{-i{\chi}({\bf r})}$. It appears due to the use of the particle number non-conserving formalism.

 \section{Josephson effect}
 \label{sec5}
 
%\begin{figure}
%\begin{center}
%\includegraphics[scale=1]{Sections/Andreev4.pdf}
%\end{center}
%\end{figure}

%\begin{figure}
%\begin{center}
%\includegraphics[scale=0.6]{Josephson.eps}
%\end{center}
%\caption{　}
%\label{Josephson}
%\end{figure}
The Josephson effect is also the effect which, in the standard theory, arises from the phase of the pair-potential \cite{Josephson62}.
We have derived the Josephson relation by the new theory, previously \cite{Koizumi2011,koizumi2020c,koizumi2020,HKoizumi2015}. We revisit here with referring to recent experimental results \cite{Bocquillon2017,Ueda2020,zhang2021}.

In considering the Josephson effect in the new theory, we note that two distinct cases arises; one is the case where the Bogoliubov excitations are common in the two superconductors, and
the other where two separate Bogoliubov excitations exist.  In the standard theory, only the latter case is considered.
We would like to emphasize that in the new theory, supercurrent is a flow of electrons by a collective mode arising from the Berry connection, not a flow of Cooper-pairs.

Let us consider a superconductor-insulator-superconductor (SIS) junction.
The hopping Hamiltonian between the two superconductors (we use labels $L$ for left and $R$ for right superconductors, respectively) across the insulator is given by
  \begin{eqnarray}
H_{LR}=-\sum_{\sigma}T_{LR} \left(  c^{\dagger}_{L \sigma} c_{R \sigma}+c^{\dagger}_{R\sigma} c_{L \sigma}  \right)
\label{juncH}
\end{eqnarray}

Let us first consider the case where the insulator part is very thin and the Bogoliubov excitations in the two superconductors are the same, which are described by the particle number conserving Bogoliubov operators $\gamma^{\dagger}_{n \sigma}$ and $\gamma_{n \sigma}$.
In this case $H_{LR}$ is re-expressed as
  \begin{eqnarray}
H_{LR}&=&-T_{L R} e^{{ i \over 2}(\hat{\chi}_L-\hat{\chi}_R)} e^{-i {e \over {\hbar c} } \int_R^L d{\bf r} \cdot {\bf A}^{\rm em}}
\nonumber
\\
&& \times
\sum_{n,m} \Big[
(
(u^{n}_{L})^{\ast}\gamma^{\dagger }_{n \downarrow} + v^{n}_{L }\gamma_{n \uparrow}) ( u^{m}_{R }\gamma_{m \downarrow}+ (v^{m}_{R})^{\ast}\gamma_{m \uparrow}^{\dagger} ) 
\nonumber
\\
&&+
(
(u^{n}_{L })^{\ast}\gamma^{\dagger }_{n \uparrow} - v^{n}_{L}\gamma_{n \downarrow}) ( u^{m}_{R }\gamma_{m \uparrow}- (v^{m}_{R})^{\ast}\gamma_{m  \downarrow}^{\dagger} )  
\Big]+\mbox{h.c.}
\end{eqnarray}

Then, the effective hopping Hamiltonian that does not cause Bogoliubov excitations is given by
\begin{eqnarray}
H_J^{e}=C \cos \left[{ {e \over {\hbar c}} \int_R^L d{\bf r} \cdot \left({\bf A}^{\rm em} -{{\hbar c} \over {2e}} \nabla \chi\right)} + \alpha \right]
\label{e}
\end{eqnarray}
where $C$ and $\alpha$ are parameters given through the following relations
\begin{eqnarray}
{ 1 \over 2} C e^{i \alpha}=-2T_{L R} \sum_{n}
 v^{n}_{L }(v^{n}_{R})^{\ast}+\mbox{h.c.}
\end{eqnarray}

Next, we consider the second case where the Bogoliubov excitations in the two superconductors are different. We denote them, $\gamma^{\dagger}_{L n \sigma}, \gamma_{L n \sigma}$ for the left superconductor, and $\gamma^{\dagger}_{R n \sigma}, \gamma_{R n \sigma}$ for the right  superconductor, respectively.

In this case $H_{LR}$ is re-expressed as
  \begin{eqnarray}
H_{LR}&=&-T_{L R} e^{ { i \over 2}(\hat{\chi}_L-\hat{\chi}_R)} e^{-i {e \over {\hbar c}} \int_R^L d{\bf r} \cdot {\bf A}^{\rm em}}
\nonumber
\\
&& \times
\sum_{n,m} \Big[
(
(u^{n}_{L})^{\ast}\gamma^{\dagger }_{L n \downarrow} + v^{n}_{L }\gamma_{L n \uparrow}) ( u^{m}_{R }\gamma_{R m \downarrow}+ (v^{m}_{R})^{\ast}\gamma_{R m \uparrow}^{\dagger} ) 
\nonumber
\\
&&+
(
(u^{n}_{L })^{\ast}\gamma^{\dagger }_{L n \uparrow} - v^{n}_{L}\gamma_{L n \downarrow}) ( u^{m}_{R }\gamma_{R m \uparrow}- (v^{m}_{R})^{\ast}\gamma_{R m  \downarrow}^{\dagger} )  
\Big]+\mbox{h.c.}
\end{eqnarray}
The current flow without Bogoliubov excitations require the second order perturbation in this case.

The second order effective Hamiltonian with taking average over the Bogoliubov excitations is given by
\begin{eqnarray}
&&\left\langle H_{LR}{1 \over {E_0 - H_0}} H_{LR} \right\rangle_{\rm Bog}
\nonumber
\\
& \approx& - \Big\langle \sum_{m, n, m', n'}T_{L R}^2 
\left[ e^{ {i \over 2} (\hat{\chi}_L-\hat{\chi}_R)}e^{-i {e \over {\hbar c}} \int_R^L d{\bf r} \cdot {\bf A}^{\rm em}}v_L^{n}u_R^{m}(\gamma_{L n \uparrow} \gamma_{R m \downarrow}-\gamma_{L n \downarrow}\gamma_{R m \uparrow})+ (L \leftrightarrow R) 
\right]
\nonumber
\\
&\times& {1 \over {\epsilon_m^{R}+\epsilon_n^{L}}}
\left[ e^{ {i \over 2} (\hat{\chi}_L-\hat{\chi}_R)}e^{-i {e \over {\hbar c}} \int_R^L d{\bf r} \cdot {\bf A}^{\rm em}}(u_{L}^{ n'}v_{R}^{ m'})^{\ast} (\gamma^{\dagger}_{L n' \downarrow} \gamma^{\dagger}_{R m' \uparrow}- \gamma^{\dagger}_{L n' \uparrow} \gamma^{\dagger}_{R m' \downarrow})+ (L \leftrightarrow R) 
\right] \Big\rangle_{\rm Bog} 
\nonumber
\\
&\approx& - \sum_{m, n} {{2T_{L R}^2 } \over {\epsilon_m^{R}+\epsilon_n^{L}}}
\Big[ v_L^{n}u_R^{m} (u_{L}^{ n}v_{R}^{ m})^{\ast}e^{ {i } (\hat{\chi}_L-\hat{\chi}_R)}e^{-i {{2e} \over {\hbar c}} \int_R^L d{\bf r} \cdot {\bf A}^{\rm em}}
\nonumber
\\
&+&(v_L^{n}u_R^{m})^{\ast} u_{L}^{ n}v_{R}^{ m}e^{ {i} (\hat{\chi}_L-\hat{\chi}_R)}e^{i {{2e} \over {\hbar c}} \int_R^L d{\bf r} \cdot {\bf A}^{\rm em}}
+|u_{L}^{ n}v_{R}^{ m}|^2+ |v_{L}^{ n}u_{R}^{ m}|^2\Big]
\label{Perttransfer3}
\end{eqnarray}

Thus, the effective hopping Hamiltonian for this case is
\begin{eqnarray}
H_J^{2e}=C' \cos \left( {{2e} \over {\hbar c} } \int_R^L d{\bf r} \cdot \left[{\bf A}^{\rm em} -{{\hbar c} \over {2e}} \nabla \chi \right] + \alpha' \right)
\label{2e}
\end{eqnarray}
where $C'$ and $\alpha'$ are parameters given through the following relations,
\begin{eqnarray}
{ 1 \over 2} C' e^{i \alpha'}=- \sum_{m, n} {{2T_{L R}^2 } \over {\epsilon_m^{R}+\epsilon_n^{L}}}(v_L^{n}u_R^{m})^{\ast} u_{L}^{ n}v_{R}^{ m}
\end{eqnarray}

This is the well-known result  in the standard theory. It gives rise to the Ambegaokar-Baratoff relation\cite{Ambegaokar} for the dc Josephson effect\cite{Josephson62}.
In this case, the current flow is the flow of electron pairs. 

Let us write the current through the junction as
\begin{eqnarray}
J=J_c \sin \phi
\label{eq90}
\end{eqnarray}

If we use $H_J^{e}$, $J_c$ is given by
\begin{eqnarray}
J_c =C{e \over {\hbar }}
\end{eqnarray}
and $\phi$ is given by
\begin{eqnarray}
\phi={ {e \over {\hbar c}} \int_R^L d{\bf r} \cdot \left({\bf A}^{\rm em} -{{\hbar c} \over {2e}} \nabla \chi\right)} + \alpha 
\end{eqnarray}

The time-derivative of $\phi$ is calculated as
\begin{eqnarray}
 \dot{\phi}={e \over {\hbar c}} \int_R^L d{\bf r} \cdot \left(\partial_t{\bf A}^{\rm em} -{{\hbar c} \over {2e}} \nabla \partial_t{\chi} \right)
 =-{e \over \hbar} \int_R^L d{\bf r} \cdot {\bf E}^{\rm em}  -\left. {e \over \hbar}  \left(  \varphi^{\rm em} +{\hbar \over {2e}}  \partial_t{\chi} \right)\right|^L_R
 \nonumber
 \\
 \label{eqnPhidot}
 \end{eqnarray}
 where
\begin{eqnarray}
 {\bf E}^{\rm em} =-{1 \over c}\partial_t{\bf A}^{\rm em} - \nabla \varphi^{\rm em}
 \end{eqnarray}
 is used.
 
 There are two contributions for $\dot{\phi}$.
 The first one is
 \begin{eqnarray}
 -{e \over \hbar} \int_R^L d{\bf r} \cdot {\bf E}^{\rm em} ={{eV} \over \hbar}
  \end{eqnarray}
 where
\begin{eqnarray}
V=- \int_R^L d{\bf r} \cdot {\bf E}^{\rm em}
  \end{eqnarray}
   is the voltage across the junction.
   
   The second contribution includes the following
     \begin{eqnarray}
  \varphi^{\rm eff}= \varphi^{\rm em} +{\hbar \over {2e}}  \partial_t{\chi} 
  \end{eqnarray}
  
  It is actually the time-component of the four vector whose spatial components are
   \begin{eqnarray}
   {\bf A}^{\rm eff}={\bf A}^{\rm em} -{{\hbar c} \over {2e}}  \nabla {\chi} 
  \end{eqnarray}

It is gauge invariant since ${\bf A}^{\rm eff}$ is gauge invariant.
 Actually, 
     \begin{eqnarray}
 \int  d^3 r \varphi^{\rm eff} \rho
   \end{eqnarray}
   appears in the Hamiltonian, thus, $\varphi^{\rm eff}$ should be related to the chemical potential $\mu$ as
       \begin{eqnarray}
 \mu=e \varphi^{\rm eff}
   \end{eqnarray}
   
 Then, Eq.~(\ref{eqnPhidot}) becomes
   \begin{eqnarray}
 \dot{\phi}={e \over \hbar} V + {1 \over \hbar}  \left( \mu_R-\mu_L \right)
 \label{eqJ1}
 \end{eqnarray}

The balance between the voltage and the chemical potential difference yields,
  \begin{eqnarray}
 eV =\mu_R-\mu_L 
 \end{eqnarray}
 
 Thus, the following Josephson relation is obtained
\begin{eqnarray}
\dot{\phi}={{2eV} \over \hbar}
\label{eqJR}
\end{eqnarray}
This leads to the Shapiro steps
\begin{eqnarray}
V_n={{h f} \over {2e}}n, \quad n :\mbox{integer}
\end{eqnarray}
    where $f$ is the microwave frequency appleid to the junction \cite{Shapiro63}. This agrees with the experimental result.

In the standard theory, $H_J^{2e}$ is used to obtain the Josephson relation without including the contribution from the chemical potential difference. However, the contribution from the chemical potential difference term arising from the electron flow from and  to the leads connected to the junction
 exists in order to maintain the chemical potentials. Thus, the omission of this term in the standard theory is not justified.
 
If this term is included, the derivation using $H_J^{2e}$ gives
\begin{eqnarray}
J_c =C'{{2e} \over {\hbar }}
\end{eqnarray}
and 
\begin{eqnarray}
\phi={ {{2e} \over {\hbar c}} \int_R^L d{\bf r} \cdot \left({\bf A}^{\rm em} -{{\hbar c} \over {2e}} \nabla \chi\right)} + \alpha' 
\end{eqnarray}
Then,  $\dot{\phi}={{4eV} \over \hbar}$ is obtained instead of Eq.~(\ref{eqJR});
in this case, the Shapiro steps become $V_n={{h f} \over {4e}}n={{h f} \over {2e}} {n \over 2}$.
Actually, the experiment that exhibits half-integer Shapiroo steps has been obtained \cite{Ueda2020}. 
This result may be attributed to the realization of $\dot{\phi}={{4eV} \over \hbar}$.

The current may be a sum of different tunneling paths. In this case, the current may be expressed as
\begin{eqnarray}
J=J_c \sum_{i_L, j_R} \sin \phi_{i_L j_R}
\end{eqnarray}
where the current is given as a sum of contributions from different paths that connect left superconductor site $i_L$ and right superconductor site $i_R$;
$J_c$ is taken to be the same irrespective of the paths for simplicity. 

Using $H_J^{e}$, the time-derivative of $\phi_{i_L j_R}$ is given by
\begin{eqnarray}
 \dot{\phi}_{i_L j_R} =-{e \over \hbar} \int_{j_R}^{i_L} d{\bf r} \cdot {\bf E}^{\rm em}  +{ 1 \over \hbar} (\mu_R - \mu_L)
  \end{eqnarray}
  where the contribution from the second term due to the chemical potential difference is independent of the sites.
  It may happen that due to the spatial and temporal fluctuations of ${\bf E}^{\rm em}$ 
  \begin{eqnarray}
  -{e \over \hbar} \int_{j_R}^{i_L} d{\bf r} \cdot {\bf E}^{\rm em}  \neq{ 1 \over \hbar} (\mu_R - \mu_L)
  \end{eqnarray}
  occurs. Actually, such fluctuations have been observed experimentally \cite{zhang2021}.
  It may also happen that $\alpha_{i_L j_R}$'s are random.
  
Taking into account the above fluctuations, the current may be given by
  \begin{eqnarray}
J&=&J_c \sum_{i_L, j_R} \sin  \left( -{e \over \hbar} \int_{j_R}^{i_L} d{\bf r} \cdot {\bf E}^{\rm em} t  +{ 1 \over \hbar} (\mu_R - \mu_L)t +\alpha_{i_L j_R}   \right) 
\nonumber
\\
&\approx&J_c   \sum_{i_L, j_R}  \left\langle \sin  \left( -{e \over \hbar} \int_{j_R}^{i_L} d{\bf r} \cdot {\bf E}^{\rm em} t +\alpha_{i_L j_R} \right) 
\right\rangle_{\rm AV}\cos  \left({ 1 \over \hbar} (\mu_R - \mu_L)t \right)
\nonumber
\\
\nonumber
&+&
J_c  \sum_{i_L, j_R}  \left\langle \cos  \left( -{e \over \hbar} \int_{j_R}^{i_L} d{\bf r} \cdot {\bf E}^{\rm em} t +\alpha_{i_L j_R}\right) \right\rangle_{\rm AV} \sin  \left({ 1 \over \hbar} (\mu_R - \mu_L)t  \right)
  \nonumber
 \\
&=&
\bar{J}_c'  \sin  \left({ 1 \over \hbar} (\mu_R - \mu_L)t  \right)+\bar{J}_c''  \cos \left({ 1 \over \hbar} (\mu_R - \mu_L)t  \right)
\end{eqnarray}
where
 \begin{eqnarray}
 \bar{J}_c'&=& J_c\sum_{i_L, j_R} 
  \left\langle \cos  \left( -{e \over \hbar} \int_{j_R}^{i_L} d{\bf r} \cdot {\bf E}^{\rm em} t +\alpha_{i_L j_R}  \right) \right\rangle_{\rm AV} 
 \nonumber
 \\
  \bar{J}_c''&=& J_c\sum_{i_L, j_R} 
  \left\langle \sin  \left( -{e \over \hbar} \int_{j_R}^{i_L} d{\bf r} \cdot {\bf E}^{\rm em} t +\alpha_{i_L j_R}  \right) \right\rangle_{\rm AV} 
\end{eqnarray}
Here, $\langle Q \rangle_{\rm AV}$ denotes the average of $Q$ taken over the paths and time-interval over the period of the applied microwave. 

In the situation where  $\bar{J}_c'$ and/or $\bar{J}_c''$ become non-zero constant, $\dot{\phi}={{eV} \over \hbar}$ will be realized.
Such a situation may arise when the above average is approximated by a gaussian distribution with mean-value zero.
In this case, the averages are calculated as
 \begin{eqnarray}
 \left\langle \sin  \left( -{e \over \hbar} \int_{j_R}^{i_L} d{\bf r} \cdot {\bf E}^{\rm em} t +\alpha_{i_L j_R}\right) 
\right\rangle_{\rm AV}&\approx&0
\nonumber
\\
 \left\langle \cos  \left( -{e \over \hbar} \int_{j_R}^{i_L} d{\bf r} \cdot {\bf E}^{\rm em} t +\alpha_{i_L j_R}  \right) \right\rangle_{\rm AV} 
  &\approx&\exp
  \left(
  -{ 1 \over 2} \left\langle \left[{ {et} \over \hbar} \int_{j_R}^{i_L} d{\bf r} \cdot {\bf E}^{\rm em}-\alpha_{i_L j_R} \right]^2 \right\rangle_{\rm AV} 
  \right)
  \nonumber
  \\
  &\approx&\exp
  \left(
  -{ 1 \over 2} \left\langle \left[{ {eV} \over {hf} }-\alpha_{i_L j_R} \right]^2 \right\rangle_{\rm AV} 
  \right)
    \nonumber
\end{eqnarray}
where the time interval is simply replaced by the microwave frequency $f^{-1}$.

Then, the Shapiro steps become $V_n={{h f} \over {e}}n={{h f} \over {2e}} {2n}$, indicating the disappearance of odd-integer Shapiro steps. Actually, such experimental results have been obtained \cite{Bocquillon2017}. In this experiment, low microwave power and low frequency cases yield disappearance of odd-integer steps; the low microwave power will result in the dominance of the 
fluctuation, and the low frequency results in the longe-time average that will cause the dominance of the fluctuation.

If the above considered situation is realized, the Shapiro steps become $V_n={{h f} \over {2e}}n$ by using $H_J^{2e}$, in agreement with the standard theory. However, the high precision of the quantized voltage seems to indicate that the result obtained by $H_J^{e}$ is the right one.

\section{Concluding remarks}
 \label{sec6}
 
 In the standard theory of superconductivity, the origin of superconductivity is the electron-pairing.
In this theory, the induced current by a magnetic field is calculated by the linear response to the vector potential, and the supercurrent is identified as the dissipationless flow of the paired-electrons, while single electrons flow with dissipation.

The above supercurrent description suffers from the following serious problems: 1) it contradicts the reversible superconducting-normal phase transition in a magnetic field observed in type I superconductors; 2) the gauge invariance of the supercurrent induced by a magnetic field requires the breakdown of the global $U(1)$ gauge invariance, or the non-conservation of the particle number; 3) the explanation of the ac Josephson effect is based on the boundary condition that is different from the real experimental one;
4) the measured London moment indicates the mass for the superconducting carrier is the free electron mass $m_e$ if the electron charge $q=-e$ is used although the standard theory predicts it to be the effective mass $m^{\ast}$.

The standard theory relies on the non-zero value of Eq.~(\ref{pair-pot}), and actually, the cause of the above problems is partly due to the belief that the non-zero value of Eq.~(\ref{pair-pot}) is a physical consequence of the Cooper instability.

In the new theory, Eq.~(\ref{pair-pot}) is replaced by Eq.~(\ref{pair-pot2}) calculated using the particle number conserving state. Non-zeroness of it is due to the presence of the Berry connection that generates the number-changing operators.
From the view point of the new theory, it appears that the standard theory takes into account the presence of the number-changing operators brought about by the Berry connection by employing the particle-number non-conserving approximation \cite{Peierls1991}. This approximation works well for some purposes. However, it also causes serious contradictions.

The new theory is a significant departure from the standard one.
We hope that the elucidation of the cuprate superconductivity is achieved with it.

\vspace{2cm}

\section*{Appendix I : Modification of Maxwell's equations in the presence of the Berry connection from many-body wave functions}

From the view point of the Feynman path integral formalism of quantum mechanics \cite{Feynmanpath}, a wave function is a sum of contributions from all paths each contributes an exponential whose phase is the classical action divided by $\hbar$ for the path in question.

For the system of charged particles and electromagnetic field, the classical action $S$ are composed of the following three terms, 
\begin{eqnarray}
S=S_1+S_2+S_3
\end{eqnarray}
where
\begin{eqnarray}
S_1=\sum_i { m \over 2} \int dt \ \dot{\bf r}_i^2
\end{eqnarray}
is the action for the particles, 
\begin{eqnarray}
S_2=-\int d^3r dt \ \left[
\rho \phi^{\rm em}({\bf r},t) -{1 \over c} {\bf j} \cdot {\bf A}^{\rm em}({\bf r},t) \right]
=-q\sum_i \int  dt \ \left[ \phi^{\rm em}({\bf r}_i,t) -{1 \over c} \dot {\bf r}_i \cdot {\bf A}^{\rm em}({\bf r}_i,t) \right]
\nonumber
\\
\end{eqnarray}
is the action for the interaction between the field and particles, and
\begin{eqnarray}
S_3={1 \over {8 \pi}}\int d^3r dt \ \left[ ({\bf E}^{\rm em})^2- ({\bf B}^{\rm em})^2 \right]
={1 \over {8 \pi}}\int d^3r dt \ \left[ \left(-\nabla \phi^{\rm em} -{1 \over c} {{\partial {\bf A}^{\rm em}} \over {\partial t}} \right)^2- \left(\nabla \times {\bf A}^{\rm em}\right)^2 \right]
\nonumber
\\
\end{eqnarray}
is the action for the field. Here $c$ is the speed of light in vacuum, and $\phi^{\rm em}$ and ${\bf A}^{\rm em}$ are the scalar and vector potentials for the electromagnetic field, respectively; $\rho$ and ${\bf j}$ are the electric charge and current densities, respectively; $q$ and $m$ are the charge and mass of the particle, respectively.

In the following we consider the case where the electric field ${\bf E}^{\rm em}$ is absent, and only the magnetic field ${\bf B}^{\rm em}$ is present.  We will consider the case  where ${\bf E}^{\rm em}$ is present, later, to deal with the ac Josephson effect.

In our previous work \cite{koizumi2019,koizumi2020c,koizumi2021}, it is shown that the Berry connection arising from many-body wave functions modifies the momentum operator $-i \hbar \nabla$ in the Schr\"{o}dinger representation of quantum mechanics as follows 
\begin{eqnarray}
-i \hbar \nabla \longrightarrow -i \hbar \nabla + \hbar {\bf A}_{\Phi}^{\rm MB}
\end{eqnarray}
where ${\bf A}_{\Phi}^{\rm MB}$ is the Berry connection defined by
 \begin{eqnarray}
{\bf A}^{\rm MB}_{\Phi}({\bf r},t)=-i \langle n_{\Phi}({\bf r},t) |\nabla|n_{\Phi}({\bf r},t) \rangle
\end{eqnarray}
and $|n_{\Phi}({\bf r}) \rangle$ is the parameterized wave function with the parameter ${\bf r}$ and integration coordinates ${\bf r}_2, \cdots {\bf r}_N$ given by
 \begin{eqnarray}
\langle  {\bf r}_{2}, \cdots, {\bf r}_{N} |n_{\Phi}({\bf r},t) \rangle = { {\Phi({\bf r}, {\bf r}_{2}, \cdots, {\bf r}_{N},t)} \over {|C_{\Phi}({\bf r} ,t)|^{{1 \over 2}}}}
\end{eqnarray}
with $|C_{\Phi}({\bf r},t)|$ being the normalization constant given by 
\begin{eqnarray}
|C_{\Phi}({\bf r},t)|=\int d{\bf r}_{2} \cdots d{\bf r}_{N}\Phi({\bf r}, {\bf r}_{2}, \cdots)\Phi^{\ast}({\bf r},  {\bf r}_{2}, \cdots)
\end{eqnarray}

 Inclusion of $\hbar {\bf A}_{\Phi}^{\rm MB}$ means the inclusion of the gauge field that describes the interaction between particles through the wave function they share.

As a consequence, the effective vector potential in the system becomes
\begin{eqnarray}
{\bf A}^{\rm eff}={\bf A}^{\rm em}+{\bf A}^{\rm fic}, \quad {\bf A}^{\rm fic}= \hbar {\bf A}_{\Phi}^{\rm MB}
\label{Aeff}
\end{eqnarray}
due to the presence of the ``fictitious'' vector potential ${\bf A}^{\rm fic}$ that arises as sthe Berry connection.

Actually, ${\bf A}^{\rm fic}$ is given by
\begin{eqnarray}
{\bf A}^{\rm fic}=-{{\hbar c} \over {2e}} \nabla \chi
\label{Afic}
\end{eqnarray}
where $\chi$ is an angular variable with period $2 \pi$ \cite{koizumi2020}. This appears through the spin-twisting itinerant motion of electrons, and the spin-twisting is caused by the Rashba spin-orbit interaction. Although the energy gain by the
 spin-twisting itself is very small, the Berry connection creates the number changing operators that make the energy gain by exploiting the
Cooper instability possible. In other words, as far as the energy gain by the electron-pair formation exceeds other energy deficits,
the spin-twisting itinerant motion of electrons occurs and non-trivial Berry connection is generated.

By including the Berry connection, $S_2$ becomes,
\begin{eqnarray}
S_2'={1 \over c}\int d^3r dt \ {\bf j} \cdot {\bf A}^{\rm eff}({\bf r},t)
\end{eqnarray}
where we only retain the term with vector potential assuming that electric field is absent.
This gives rise to ``Lorentz force" in the classical electromagnetic dynamics. The Lorenz force from ${\bf A}^{\rm fic}$ is zero in classical mechanics; however, 
${\bf A}^{\rm fic}$ may affect the dynamics of charged particles through the Aharonov-Bohm effect \cite{AB1959} in quantum mechanics.
Actually, this effect is the main concern of the present work. We call this term, the Lorentz interaction term, instead of the Lorentz force term,
to emphasize it contains the  Aharonov-Bohm effect.

Since the electromagnetic field energy is the energy stored in the space through the Lorentz interaction term $S_2'$,
$S_3$ should be modified using ${\bf B}^{\rm eff}={\bf B}^{\rm em}+{\bf B}^{\rm fic}$, 
\begin{eqnarray}
S_3'=-{1 \over {8 \pi}}\int d^3r dt \ ({\bf B}^{\rm eff})^2 
\end{eqnarray}
where 
\begin{eqnarray}
 {\bf B}^{\rm fic}=\nabla \times {\bf A}^{\rm fic}={{\hbar c} \over {2e}} \nabla \times \nabla \chi
\end{eqnarray}

${\bf B}^{\rm fic}$ may not be zero due to the fact that $\chi$ may be multi-valued.

Using $S_2'$ and $S_3'$, two of the Maxwell's equations are modified as
\begin{eqnarray}
\nabla \cdot {\bf B}^{\rm eff}&=&0
\label{Maxwell1}
\\
\nabla \times {\bf B}^{\rm eff}&=&{{4 \pi } \over c}{\bf j}
\label{mMax2}
\end{eqnarray}

The first one gives rise to a Dirac monopole as shown below. It is written as
\begin{eqnarray}
\nabla \cdot {\bf B}^{\rm em}=-\nabla \cdot (\nabla \times {\bf A}^{\rm fic})
\end{eqnarray}
When the both sides of the above equation are integrated for a closed region with surface ${\rm Sf}$,
we have
\begin{eqnarray}
\int_{\rm Sf} d{\bf S} \cdot {\bf B}^{\rm em}=-\int_{\rm Sf} d{\bf S} \cdot (\nabla \times {\bf A}^{\rm fic})
\end{eqnarray}

We split ${\rm Sf}$ into two surfaces ${\rm Sf}_1$ and ${\rm Sf}_2$ with common boundary loop, $C=\partial ({\rm Sf}_1)=-\partial ({\rm Sf}_2)$. Then, we have 
\begin{eqnarray}
\int_{{\rm Sf}_1} d{\bf S} \cdot (\nabla \times {\bf A}^{\rm fic})+\int_{{\rm Sf}_2} d{\bf S} \cdot (\nabla \times {\bf A}^{\rm fic})
=\int_{\partial( {\rm Sf}_1)} d{\bf r} \cdot {\bf A}^{\rm fic}+\int_{\partial( {\rm Sf}_1)} d{\bf r} \cdot {\bf A}^{\rm fic}
\nonumber
\\
\end{eqnarray}

We examine the case in which singularities exist in ${\bf A}^{\rm fic}$. 
Let us consider a closed surface $S$ with boundary $C=\partial S$, and ${\bf A}^{\rm fic}$ has a singularity in $S$.
Then, we have
\begin{eqnarray}
\int_{C} d{\bf r} \cdot {{\hbar c} \over {2e}} \nabla \chi ={{h c} \over {2e}} n 
\end{eqnarray}
where $n$ is an integer. 

If we have $n=0$ for $\partial( {\rm Sf}_1)$ term, and $n=1$ for $\partial( {\rm Sf}_2)$ term, we have
\begin{eqnarray}
\int_{\rm Sf} d{\bf S} \cdot {\bf B}^{\rm em}={{h c} \over {2e}} 
\end{eqnarray}
This shows that a monopole with magnetic charge ${{h c} \over {2e}}$ exists in the region enclosed by ${\rm Sf}$. This corresponds to the monopole considered by Dirac \cite{Monopole}.

The second one in Eq.~(\ref{mMax2}) is equal to 
\begin{eqnarray}
\nabla \times {\bf B}^{\rm em}&=&{{4 \pi } \over c}{\bf j}
\label{Maxwell2}
\end{eqnarray}
since $\nabla \times {\bf B}^{\rm fic}=0$ is satisfied as shown below;
it is well-known that $\nabla \chi$ in ${\bf A}^{\rm fic}$ can be decomposed as
\begin{eqnarray}
\nabla \chi=\nabla \chi_0 +\nabla f, \quad \nabla^2 \chi_0=0
\end{eqnarray}
where $f$ is a single-valued, and $\chi_0$ may be multi-valued. Thus, we have 
\begin{eqnarray}
\nabla \times {\bf B}^{\rm fic}=\nabla \times (\nabla \times \nabla \chi_0)= \nabla (\nabla^2 \chi_0)-\nabla^2 \nabla \chi_0=0
\label{Bfic0}
\end{eqnarray}

As a consequence, Eq.~(\ref{mMax2}) is reduced to the original one in Eq.~(\ref{Maxwell2}).

\section*{Appendix II : Modification of the Ginzburg-Landau theory including the Berry connection from many-body wave functions and its consequences}

 The Ginzburg-Landau theory \cite{GL} is based on the London theory \cite{London1950}. In the London theory, the velocity field for electrons in superconductors is given by
\begin{eqnarray}
 {\bf v}=-{ q \over {mc}} \left( {\bf A}^{\rm em}  - { {c \hbar} \over q} \nabla \chi^{\rm super} \right)
\label{eqr1}
\end{eqnarray}
where $\chi^{\rm super}$ is the superpotential assumed to exist in superconductors.

The Ginzburg-Landau theory uses a free energy consists of the material part and the magnetic field part. It assumes the presence of the effective wave function of superconducting electrons $\Psi_{\rm GL}$ in the superconducting phase. 

Using $\Psi_{\rm GL}$, the material part of the free energy for a superconductor is given by
 \begin{eqnarray}
 F_{\rm mat}=F_{\rm normal}+
 \int d^3 r {1 \over {2m}}  \left| \left( { \hbar \over i} \nabla -{q \over c}{\bf A}^{\rm em} \right) \Psi_{\rm GL} \right|^2
 +  \int d^3 r  \left( \alpha  |\Psi_{\rm GL}|^2 +{ \beta \over 2}|\Psi_{\rm GL}|^4 \right)
 \nonumber
 \\
 \label{mat1}
\end{eqnarray}
where $\alpha $ is a negative real parameter, and $\beta$ is a positive real parameter..

We can express $\Psi_{\rm GL}$ using the supercurrent carrier density $n_s$ and the superpotential $\chi^{\rm super}$ as
 \begin{eqnarray}
 \Psi_{\rm GL}=n_s^{1/2}e^{ i \chi^{\rm super}}
 \end{eqnarray}
 
 Then, the kinetic term becomes
  \begin{eqnarray}
 \int d^3 r {1 \over {2m}}  \left| \left( { \hbar \over i} \nabla -{q \over c}{\bf A}^{\rm em} \right) \Psi_{\rm GL} \right|^2
 =F_k + \int d^3 r {{\hbar^2 (\nabla n_s)^2} \over {8m \ n_s}} 
  \label{mat2}
\end{eqnarray}
where the supercurrent kinetic energy is given by
\begin{eqnarray}
 F_k=\int d^3 r {m \over {2}}n_s{\bf v}^2=\int d^3 r {{q^2 n_s} \over {2mc^2}} \left( {\bf A}^{\rm eff} \right)^2
\end{eqnarray}
Here, ${\bf A}^{\rm eff}$ in Eq.~(\ref{Aeff}) is used by identifying
\begin{eqnarray}
 {\bf A}^{\rm fic} =- { {c \hbar} \over q} \nabla \chi^{\rm super}=- { {c \hbar} \over {2e}} \nabla \chi, 
\label{eqr3}
\end{eqnarray}
assuming that $\chi^{\rm super}$ arises from the Berry connection.

In the standard theory, the Ginzburg-Landau theory is derived from the BCS theory,
yielding $q=-2e$ \cite{Gorkov1959}. In this case the mass of the charge carriers becomes $m=2m^{\ast}$, thus,
${ m \over q}$ in the London moment becomes $-{ m^{\ast} \over e}$, disagrees with the experimental value $-{ m_e \over e}$.
This indicates that the accepted derivation of the Ginzburg-Landau theory from the standard theory is incorrect.

We will use $q=-e$ here, and identify
\begin{eqnarray}
 \nabla \chi^{\rm super}= -{1 \over 2 }\nabla \chi
\end{eqnarray}
with $m=m_e$. This relation, $m=m_e$, can be explained in the new theory \cite{koizumi2021}.

Using  $F_k$, the current density ${\bf j}$ is calculated as
\begin{eqnarray}
{\bf j}=-c {{\delta F_k} \over {\delta {{\bf A}^{\rm em}}}}=-{{e^2 n_s} \over {m_e c}} {\bf A}^{\rm eff}
\label{j-Aeff}
\end{eqnarray}

Since
\begin{eqnarray}
{\bf j}=-e n_s{\bf v}
\end{eqnarray}
Eq.~(\ref{j-Aeff}) is equivalent to the London equation in Eq.~(\ref{eqr1}).
Actually, the above relation should be regarded as the definition of $n_s$ through ${\bf j}$ and ${\bf v}$. 

Now, consider the magnetic field part of the GL free energy,
\begin{eqnarray}
 F_m=\int d^3 r {1 \over {8\pi}} \left( {\bf B}^{\rm eff}  \right)^2 
 \end{eqnarray}
This is different from the one employed by the original GL work due to the use of ${\bf B}^{\rm eff}$ in place of ${\bf B}^{\rm em}$. 
 
 The stationary condition of $F_k+F_m$ with respect to the variation of ${\bf A}^{\rm fic}$ yields,
   \begin{eqnarray}
 -{1 \over c}{\bf j}+{ 1 \over {4 \pi}}\nabla \times {\bf B}^{\rm eff}= -{1 \over c}{\bf j}+{ 1 \over {4 \pi}}\nabla \times {\bf B}^{\rm em}=0
\label{Maxellj-B}
 \end{eqnarray}
 where Eq.~(\ref{Bfic0}) is used.
This is one of the Maxwell's equations.

Using Eq.~(\ref{j-Aeff}) and neglecting the spatial variation of $n_s$, we have
 \begin{eqnarray}
\nabla \times {\bf j}=-{{ n_s e^2} \over {m_e}}\left[\nabla \times{\bf A}^{\rm em} +\nabla \times {\bf A}^{\rm fic} \right]=
-{{ n_s e^2} \over {m_e}}\left[{\bf B}^{\rm em} +\nabla \times {\bf A}^{\rm fic} \right]
\label{E1}
\end{eqnarray}

From Eq.~(\ref{Maxellj-B}), the following relation is obtained,
  \begin{eqnarray}
\nabla \times {\bf j}={{ c} \over {4 \pi }}\nabla \times (\nabla \times {\bf B}^{\rm em})= -{{ c} \over {4 \pi }}\nabla^2  {\bf B}^{\rm em}
\label{E2}
\end{eqnarray}
Here, $\nabla \cdot {\bf B}^{\rm em}=0$ is assumed.

Combining Eqs.~(\ref{E1}) and (\ref{E2}), the following is obtained,
 \begin{eqnarray}
\nabla^2 {\bf B}^{\rm em}-{{1} \over {\lambda^2}}{\bf B}^{\rm em}={{1} \over {\lambda^2}}\nabla \times {\bf A}^{\rm fic} ={{1} \over {\lambda^2}}{\bf B}^{\rm fic} 
\label{Abr}
\end{eqnarray}
where $\lambda$ is the London penetration depth 
 \begin{eqnarray}
\lambda =  \sqrt{ {m_e c} \over { 4 \pi n_s e^2}}
\end{eqnarray}

Now we consider the loop current formation by following Abrikosov \cite{Abrikosov}. Actually, the presence of ${\bf B}^{\rm fic}=\nabla \times {\bf A}^{\rm fic}$ in Eq.~(\ref{Abr}) naturally gives rise to vortices. 
The characteristic length scale for the spatial variation of $n_s$ in the Ginzburg-Landau theory is 
   \begin{eqnarray}
\xi_{\rm GL}=\sqrt{ \hbar^2 \over {2m_e|\alpha|}}
\end{eqnarray}

Abrikosov argued that if  $\lambda \gg \xi_{\rm GL}$ is satisfied and the singularity of $\nabla \chi$ is along the $z$-axis, 
Eq.~(\ref{Abr}) can be approximated as
 \begin{eqnarray}
\nabla^2 {\bf B}^{\rm em}-{{1} \over {\lambda^2}}{\bf B}^{\rm em}={{1} \over {\lambda^2}} \Phi^{\rm fic} {\bf e}_z \delta^{(2)}({\bf r})
\end{eqnarray}
in the region away from the core, 
 where $\Phi^{\rm fic}$ is given by
 \begin{eqnarray}
 \Phi^{\rm fic} =\int_{S} {\bf B}^{\rm fic} \cdot d{ \bf S} = \int_{C} {\bf A}^{\rm fic} \cdot d{ \bf r}=
  { {c \hbar} \over {2e}} \int_{C} \nabla \chi \cdot d{ \bf r}
 \end{eqnarray}
   and $\delta^{(2)}({\bf r})$ is the delta function in two-dimension with singularities along the $z$-axis.
   
 The solution is known to be ${\bf B}^{\rm em}={ B}^{\rm em}(r){\bf e}_z$, where ${ B}^{\rm em}(r)$ is given by
  \begin{eqnarray}
 { B}^{\rm em}(r)= -2\pi \Phi^{\rm fic}\lambda^{-2}K_0(r/\lambda)
  \end{eqnarray}
  Here $K_0$ is the modified Bessel function of the 2nd kind, and $r$ is the distance from the $z$-axis.
  This expresses a vortex along the $z$-axis with core size $\xi_{\rm GL}$, accompanied by loop current around it.
  Thus, $\xi_{\rm GL}$ can be identified as the core size of the loop current that exits in a superconductor.
  
  In the BCS theory, a different coherence length 
    \begin{eqnarray}
 \xi_{\rm BCS}= {{ \hbar v_{\rm Fermi}} \over {\pi \Delta}}
 \label{xi-BCS}.
  \end{eqnarray}
  is defined, where $ v_{\rm Fermi}$ is the velocity of the electron at the Fermi energy.  
  
  It is known that $\xi_{\rm GL}$ and $\xi_{\rm BCS}$ are similar in size at very low temperatures for BCS superconductors.
  However, $\xi_{\rm BCS}$ is regarded as the size of the Cooper pair.  
  In our previous work, it has been argued that the Cooper pair formation is accompanied by the loop current formation that
  encircles a section of the Fermi surface. This loop current gives rise to ${\bf A}^{\rm fic}$
  given in Eq.~(\ref{Afic}) \cite{koizumi2020,koizumi2020c}. 
  Actually, we can relate $\xi_{\rm BCS}$ to the core size of this loop current as explained below.
  
  First, we associate $\xi_{\rm BCS}$ to the wave number $q_c$ that has an excitation energy equal to the gap energy $\Delta$,
   \begin{eqnarray}
\Delta=\hbar q_c v_{\rm Fermi}
  \end{eqnarray}
  
  If we identify 
     \begin{eqnarray}
 \xi_{\rm BCS} ={ 1 \over {\pi q_c}}
  \end{eqnarray}
  we obtain Eq.~(\ref{xi-BCS}). This may be interpreted that $\xi_{\rm BCS}$ is an estimate of the size of the loop current whose excitation energy is equal to the gap energy. 
  
  The presence of the loop current by ${\bf A}^{\rm fic}$ is plausible from the experimental fact
  that the superconducting - normal metal phase transitions in a magnetic field are reversible \cite{koizumi2020b}
  since it explains the 
  the reversible energy transfer between the kinetic energy of the supercurrent and the magnetic field energy.
  We will explain this point below: by taking into account only the change of ${\bf A}^{\rm fic}$ in the time interval $\Delta t$ during the phase transition, the change of the kinetic energy is given by
  \begin{eqnarray}
 \Delta F_k&=&\int d^3 r {{e^2 n_s} \over {m_e c^2}}  {\bf A}^{\rm eff} \cdot \int_t^{t+\Delta t} \partial_t {\bf A}^{\rm fic}dt
 \nonumber
 \\
 &=&-{1 \over c} \int d^3 r {\bf j}\cdot \int_t^{t+\Delta t} \partial_t {\bf A}^{\rm fic}dt
\end{eqnarray}
  and the change of the magnetic field energy is given by
  \begin{eqnarray}
 \Delta F_m&=&\int d^3 r {1 \over {4 \pi}} {\bf B}^{\rm eff} \cdot \int_t^{t+\Delta t} \partial_t {\bf B}^{\rm fic}dt
 \nonumber
 \\
 &=&\int d^3 r {1 \over {4 \pi}}\nabla \times {\bf B}^{\rm eff} \cdot \int_t^{t+\Delta t} \partial_t {\bf A}^{\rm fic}dt
  \nonumber
 \\
 &=&{1 \over c} \int d^3 r {\bf j} \cdot \int_t^{t+\Delta t} \partial_t {\bf A}^{\rm fic}dt
 \end{eqnarray}
 Thus, the free energy conservation, $\Delta F_m+ \Delta F_k=0$, is satisfied \cite{koizumi2020b}.
 
 The transition with $ \Delta F_{\rm mag}+ \Delta F_{\rm kin}=0$ in the time interval $\Delta t$ occurs via 
 \begin{eqnarray}
 \int_t^{t+\Delta t} dt \ \partial_t {\bf A}^{\rm fic} 
\end{eqnarray}
This gives rise to a quantized change without Joule heating brought about by the modification of the winding numbers for $\chi$. 
This is the key step to realize the reversible superconducting-normal metal phase transition in a magnetic field.
Other changes occur among ${\bf j}_s$, ${\bf B}^{\rm em}$, $n_s$, 
to satisfy $\nabla {\bf B}^{\rm em}={ {4\pi} \over c}{\bf j}_s$ and $\partial_t n_s +\nabla \cdot {\bf j}_s=0$.
However, those changes can proceed without Joule heating \cite{koizumi2020b}. 

If we assume that the situation considered above is smoothly connected to the ${\bf B}^{\rm em}=0$ case, the existence of the loop currents with net zero macroscopic current is expected. 
 In other word, the phase transition between the superconducting and normal phases occurs through the creation and annihilation of the loop currents 
 generated by ${\bf A}^{\rm em}$ in general.
 
Let us consider the free energy balance for the ${\bf B}^{\rm em}=0$ case.
In this case ${\bf A}^{\rm em}$ is a pure gauge, and it cancels ${\bf A}^{\rm fic}$ except in the core region of the size $\xi$. 

For simplicity, we assume that singularities of $\chi$ form vortices along the $z$-direction. 
 Let us pick up one of them, and take the $z$-axis along it. 
 Then, the supercurrent carrier density is given by
 \begin{eqnarray}
n_s=n_0 e^{ -{{2 r} \over {\xi}}}
\end{eqnarray}
near the vortex,  where $n_0$ is a constant.
 
The sum of the energies from the spatial variation of $n_s$ and the term linear to $n_s$ in Eqs.~(\ref{mat1}) and (\ref{mat2}) is given by
   \begin{eqnarray}
 \int d^3 r {{\hbar^2 (\nabla n_s)^2} \over {8m_e \ n_s}} +\int d^3 r  \alpha  |\Psi_{\rm GL}|^2 =
  \int d^3 r {{\hbar^2 (\nabla n_s)^2} \over {8m_e \ n_s}} +\int d^3 r  \alpha  n_s
\end{eqnarray}

It becomes zero when $\xi$ satisfies 
   \begin{eqnarray}
\xi=\xi_{\rm GL}
\end{eqnarray}
If $\xi > \xi_{\rm GL}$, it is negative, indicating that the vortex formation may be possible if the energy gain from it is more than the energy deficit from the core formation.
In other words, if the energy deficit by the creation of vortex cores is compensated by the energy gain by the generation of the non-trivial Berry connection, the loop currents generation by the Berry connection will be realized.

The energy gain here comes from the electron-pair formation in the BCS superconductors.
In the new theory, the non-trivial Berry connection ${\bf A}^{\rm fic}$ appears when the spin-twisting itinerant motion of the core size $\xi_{\rm BCS}$ is realized, leading to the relation $\xi_{\rm GL} \approx \xi_{\rm BCS}$.  
 The non-trivial Berry connection ${\bf A}^{\rm fic}$ and the pairing energy gap appear, simultaneously; thus, the pairing energy gap can be used as the superconducting order parameter.

% BibTeX users please use one of
%\bibliographystyle{spbasic}      % basic style, author-year citations
%\bibliographystyle{spmpsci}      % mathematics and physical sciences
\bibliographystyle{spphys}       % APS-like style for physics
%\bibliography{}   % name your BibTeX data base

%\bibliography{SPIN-BCS}

\end{document}